\documentclass[twocolumn,showpacs]{revtex4}
\usepackage{amsmath,amssymb,graphicx,bm}
\newcommand{\be}{\begin{equation}}
\newcommand{\ee}{\end{equation}}
\newcommand{\U}{\widetilde{U}}
\newcommand{\ra}{\rangle}
\newcommand{\la}{\langle}

\begin{document}

\title{Quantum chaotic resonances from short periodic orbits}

\author{M. Novaes$^{1,2}$,  J.M. Pedrosa$^3$, D. Wisniacki$^4$, G.G. Carlo$^3$ and J.P. Keating$^1$}

\affiliation{$^1$School of Mathematics, University of Bristol,
Bristol BS8 1TW, UK\\$^2$Departamento de F\'isica, Universidade
Federal de S\~ao Carlos, S\~ao Carlos, SP, 13565-905,
Brazil\\$^3$Departamento de F\'isica, CNEA, Av. Libertador 8250,
Buenos Aires C1429BNP, Argentina\\$^4$Departamento de F\'isica,
FCEyN, UBA, Ciudad Universitaria, Buenos Aires C1428EGA, Argentina}


\begin{abstract}
We present an approach to calculating the quantum resonances and
resonance wave functions of chaotic scattering systems, based on the
construction of states localized on classical periodic orbits and
adapted to the dynamics. Typically only a few of such states are
necessary for constructing a resonance. Using only short orbits
(with periods up to the Ehrenfest time), we obtain approximations to
the longest living states, avoiding computation of the background of
short living states. This makes our approach considerably more
efficient than previous ones.  The number of long lived states
produced within our formulation is in agreement with the fractal
Weyl law conjectured recently in this setting. We confirm the
accuracy of the approximations using the open quantum baker map as
an example.
\end{abstract}

\maketitle

Quantum scattering in chaotic systems is a very active field in the area
of quantum chaos, with current experimental realizations in
ballistic quantum dots \cite{dots}, microlasers \cite{lasers} and
microwave cavities \cite{microwave}, among others \cite{class}. The
fractal Weyl law \cite{prl91wl2003}, which
relates the counting of resonances in the complex plane to the
dimension of the trapped set of the corresponding classical
dynamics, has attracted considerable attention
\cite{henning,jpa38sn2005,walsh,shepe} because resonances (or Gamow
states) are central to the description of many
aspects of wave scattering.

The decaying eigenstates associated with these quantum chaotic resonances
are far from being fully understood. They have
recently been shown \cite{casati,walsh} to display fractal structures in
phase space when the resonance is long-lived, that is when the decay
rate $\Gamma/\hbar$ remains finite as $\hbar\to0$, and to be
localized when the resonance is short-lived, that is when
$\hbar/\Gamma\to 0$. However, the semiclassical limit is much richer
for scattering systems than for closed ones, for which the
quantum ergodicity theorem \cite{ergod} states that almost all
states become uniform. Owing to the existence of different decay
rates, nothing of this kind is available for scattering systems and we are
still far from a complete description.

Our purpose here is to establish an approach to resonances and
resonance wave functions based on short classical periodic orbits.
The idea is to use the proliferation of periodic orbits in the phase
space of chaotic systems to build an approximate basis of functions
for the quantum Hilbert space. These functions are constructed in
such a way as to contain dynamical information up to Ehrenfest time.
This formulation has several virtues. First, the fractal Weyl law
emerges very naturally from the theory and is seen to have a direct
connection with periodic orbits. Second, we have an approximation to
the quantum propagator that provides the long-lived states (which
are usually dominant) without having to calculate short-lived
states, therefore very significantly reducing the dimension of the
matrices involved in the theory. Specifically, we achieve a power
saving in the matrix dimension.  Third, it turns out that usually
only a few of our states are required to produce a quantum
resonance, providing a way to quantitatively analyze scarring
effects (anomalous localization of chaotic eigenstates around
periodic orbits \cite{scars}). Finally, it opens a new and promising
avenue for semiclassical approaches to resonance wave functions,
which have so far been elusive.

A corresponding theory exists for closed systems in the form of {\it
scar functions} \cite{scarfunc,stadium}, which have proved efficient
in providing semiclassical approximations for quantum spectra and
eigenstates of billiards \cite{stadium} and quantum maps
\cite{saraceno}.  In the open systems considered here, the
efficiency gain is considerably greater. An alternative periodic
orbit approach to resonances already exists in terms of the
semiclassical zeta function \cite{zeta}. However, the orbits used
are in general much longer than the ones considered here and the
final result is an approximation to the spectral determinant that
does not provide the wave functions.

For simplicity, we restrict ourselves to quantum maps, in which
time evolution is discrete, the quantum Hilbert space has finite
dimension $N=1/(2\pi\hbar)$ and the classical phase space is a
torus. Open maps are defined by identifying a region of phase space
-usually a strip of width $M/N$ parallel to one of the axis- with a
`hole', so that particles falling into that region are lost. This is
a simplified but effective model for chaotic cavities with leads
like the ones used in experiments with quantum dots. Quantum
mechanically, the introduction of the hole corresponds to setting $M$
rows (or columns) in the quantum propagator $U$ to zero. Since it is
no longer unitary, the new matrix $\U$ has left and right eigenstates
\be\label{leftright} \U|\Psi^R_n\ra=z_n|\Psi^R_n\ra, \quad \la
\Psi_n^L|\U=z_n\la \Psi_n^L| \ee and we may choose the following
normalization and orthogonality conditions: \be\label{ortho}\la
\Psi_n^R|\Psi^R_n\ra=\la \Psi_n^L|\Psi^L_n\ra,\quad\la
\Psi_n^L|\Psi^R_m\ra=\delta_{nm}.\ee The eigenvalues lie in the unit
disk, $|z_n|^2=e^{-\Gamma_n}\leq 1$, and the quantity $\Gamma_n\geq
0$ is interpreted as the decay rate. It was shown in \cite{walsh}
that in the semiclassical limit the long-lived left and right
eigenstates localize on the stable and unstable manifolds,
respectively, of the classical trapped set (strange repeller).

\begin{figure}[t]
\includegraphics[clip,scale=0.35]{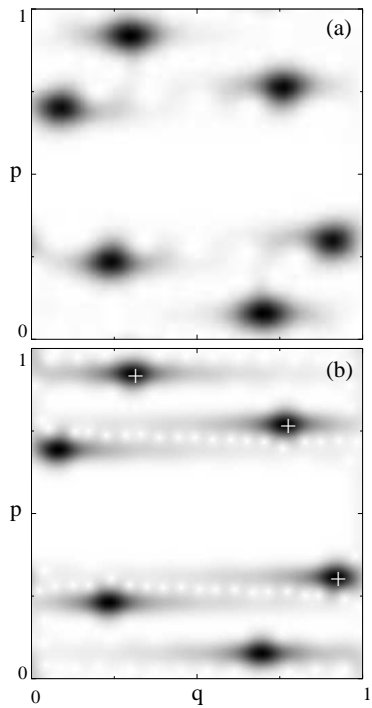}
\caption{Husimi representation of symmetrized right scar functions
corresponding to a period $3$ orbit of the triadic baker map, at
$N=81$ (a) and $N=243$ (b). In panel (b) white crosses show the
location of periodic points of the trajectory, and reflection by the
diagonal produces a symmetric partner. Absolute value grows from
white to black. \label{figfscars1}}
\end{figure}

We associate with every primitive periodic orbit $\gamma$ of
period $L$ a total of $L$ scar functions, as was done in
\cite{saraceno} for closed maps. One starts by building what is
called a tube function, or periodic orbit mode, \be
|\phi_\gamma^k\ra=\frac{1}{\sqrt{L}}\sum_{j=0}^{L-1} \exp\{-2\pi i(j
A^k_\gamma-N\theta_j)\}|q_j,p_j\ra,\ee which is a linear combination
of coherent states on each of the $L$ distinct points $(q_j,p_j)$ of
the orbit. Here $\theta_j=\sum_{l=0}^j S_l$ where $S_l$ is the
action acquired by the $l$th coherent state in one step of the map.
The total action of the orbit is $\theta_L\equiv S_\gamma$. The
quantity $A^k_\gamma=(NS_\gamma+k)/L$ is a Bohr-Sommerfeld-like
eigenvalue, \be U|\phi_\gamma^k\ra\approx\frac{e^{2\pi
iA^k_\gamma}}{\sqrt{\cosh \lambda}}|\phi_\gamma^k\ra.\ee We denote
by $\lambda$ the Lyapounov exponent of the system.

\begin{figure}[t]
\includegraphics[clip,scale=0.45]{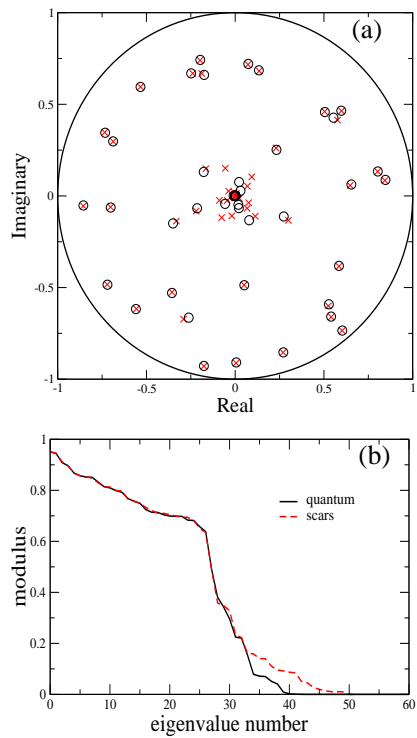}
\caption{In (a) we show the spectrum of the open triadic baker map
for $N=81$ (circles) and the spectrum of the scar matrix (crosses).
Their moduli (ordered by decreasing value) are displayed in panel
(b). \label{figspectrum}}
\end{figure}

The right and left scar functions associated with the periodic orbit
are defined through the propagation, under the open map, of the
tubes until around the Ehrenfest time $T_E=\frac{1}{\lambda} \ln N$.
Namely, \be\label{prop}
|\psi^R_{\gamma,k}\ra=\frac{1}{\mathcal{N}_\gamma^R}\sum_{t=0}^{\tau}
\U^te^{-2\pi iA^k_\gamma t}\cos\left(\frac{\pi
t}{2\tau}\right)|\phi_\gamma^k\rangle,\ee and \be
\la\psi^L_{\gamma,k}|=\frac{1}{\mathcal{N}_\gamma^L}\sum_{t=0}^{\tau}
\la\phi_\gamma^k|\U^te^{-2\pi iA^k_\gamma t}\cos\left(\frac{\pi
t}{2\tau}\right).\ee The constants $\mathcal{N}^{R,L}$ are chosen
such that $\la \psi_{\gamma,k}^R|\psi^R_{\gamma,k}\ra=\la
\psi_{\gamma,k}^L|\psi^L_{\gamma,k}\ra$ and $\la
\psi_{\gamma,k}^L|\psi^R_{\gamma,k}\ra=1$. The cosine is used to
introduce a smooth cutoff, and the propagation time $\tau$ is taken
of the order of $T_E$. In contrast to what is done for closed
systems, we do not use negative powers of the matrix $\U$, i.e. the
tubes are propagated in only one direction in time. This implies
that the phase space support of right and left scar functions
becomes localized on the unstable and stable manifolds,
respectively, of the periodic orbit, in consonance with the
properties of resonances.  It is natural to order these resonant
scar functions according to the modulus of $\la
\psi_{\gamma,k}^L|\U|\psi_{\gamma,k}^R\ra$ so that longest-living
ones come first. Finally, we note that it is convenient to impose
the symmetries of the map on these functions (if two orbits are
related by symmetry, we build symmetric and antisymmetric scar
functions).

It is by now established \cite{scarfunc, stadium, saraceno} that it
is only necessary to use short periodic orbits to obtain good
approximations to quantum spectra and eigenstates. By `short' we
mean orbits with periods up to around the Ehrenfest time of the
system. This is not unexpected, because $T_E$ is the time when
quantum interference effects become important. In the present case,
since all periodic points are on the trapped set, the theory
approximates only the long-lived states. Using the ordering
mentioned above we construct the matrix $\langle
\psi_n^L|\U|\psi_m^R\rangle$, which we call the {\it scar matrix},
as an approximation to the `long-lived sector' of $\U$. This matrix
is by construction almost diagonal (in the sense that it is equal to
a diagonal matrix plus a sparse one).

What is the dimension of the scar matrix? For chaotic systems the
number of periodic points grows with the period $L$ like $e^{hL}$
where $h$ is the topological entropy. Taking orbits with periods up
to $T_E$ we have $e^{hT_E}$ periodic points and corresponding scar
functions. However, for small openings $h$ is related to the fractal
(information) dimension of the trapped set by $d=2h/\lambda$
\cite{dimen,beck}. Since $e^{\lambda T_E}=N$ we conclude that the
matrix dimension (and the number of long-lived states) scales with
$N$ as $N^{d/2}$, in agreement with the fractal Weyl law
\cite{prl91wl2003,jpa38sn2005}.  Note that the dimension of $\U$ is
$N$, so our approach leads to a power saving in the size of the
matrices used.  This therefore represents a considerable improvement
in efficiency.

\begin{figure}[t]
\includegraphics[clip,scale=0.35]{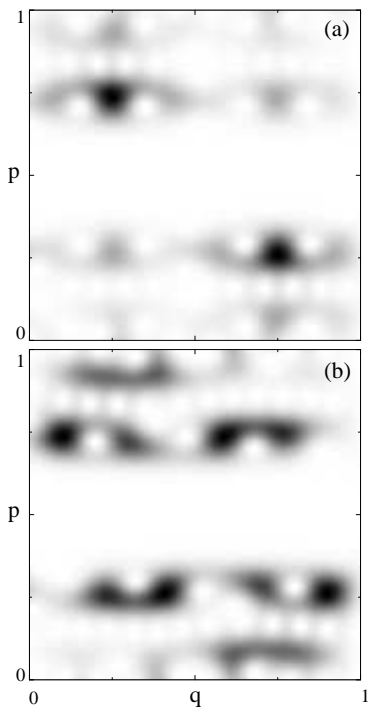}
\caption{Husimi representation of right resonances number $3$ and
$12$ (ordered by decreasing eigenvalue moduli) corresponding to the
triadic baker map at $N=81$.} \label{figresonances}
\end{figure}

The above reasoning is in a sense complementary to the one presented
in \cite{henning}. There the authors considered quantum states which
escape from the system before the Ehrenfest time (short-lived
states). As a consequence they were led to regions of phase space
that are preimages of the hole. Conversely, we are attempting to
construct the quantum states which do not escape from the system
before the Ehrenfest time (long-lived states) and are thus lead to
short periodic orbits on the repeller. Consistently, both approaches
result in the fractal Weyl law.

\begin{figure}[t]
\includegraphics[clip,scale=0.35]{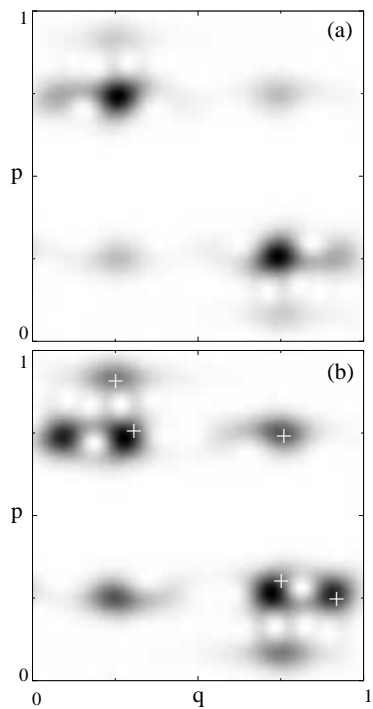}
\caption{Husimi representation of two symmetrized right scar
functions corresponding to the same orbit of period $5$, at $N=81$.
White crosses show the periodic points of the orbit, and reflection
through the diagonal produces a symmetric partner. \label{figfscars2}}
\end{figure}

As an example of the formalism, we use the triadic baker
map, as in \cite{walsh}. For this map the Lyapounov exponent is
$\lambda=\ln 3$, and we choose the quantum dimension to be $N=3^k$
so that $T_E=k$. The trapped set is the cartesian
product $Can\times Can$ where $Can$ is the usual middle-third Cantor
set of dimension $\ln 2/\ln 3$. The fractal Weyl law therefore
predicts that the number of long-lived states should grow like
$N^{\ln2/\ln3}=2^k$ (this is actually the exact number if Walsh
quantization \cite{walsh} is used). Let us take for instance $k=4$
and build scar functions for orbits with period up to $5$ (there are
$51$ periodic points in total). We illustrate this construction in
Figure \ref{figfscars1}, where we show the Husimi plots of a
symmetrized right scar function corresponding to an orbit of period
$3$, at $N=81$ and $N=243$. It can clearly be seen that the
probability extends along the unstable manifolds of this periodic
orbit ($q$ axis direction). We have used $\tau=T_E$.

In Figure \ref{figspectrum} we present the exact quantum spectrum
and the spectrum of the scar matrix (solution of a generalized
eigenvalue problem since $\la \psi_n^L|\psi_m^R\ra\neq
\delta_{nm}$), both for $N=81$. We see that the latter provides
excellent approximations to the first 30 resonances: they are all
reproduced accurately and moreover there are no spurious eigenvalues
among them. We have verified that for $N=3^5$ and using orbits up to
period $6$ (matrix dimension 106) the first 55 resonances are
reproduced accurately and without spurious eigenvalues. We have also
verified that the method works well for other baker maps.

Figure \ref{figresonances} shows the Husimi functions of the right
resonances number $3$ and $12$ (ordered according to decreasing
moduli of eigenvalues), again for $N=81$. They are both strongly
localized around periodic orbits, i.e. they are scarred. The
corresponding eigenstates of the scar matrix are indistinguishable
from the exact ones, showing that this matrix is indeed a good
approximation to the long-lived sector of $\U$.

In Figure \ref{figfscars2} we show two symmetrized scar functions
built from the same periodic orbit, of period 5. Because some of the
periodic points are very near the symmetry lines in phase space (the
diagonals of the square), we see that interference between the orbit
and its symmetric partner makes the functions look rather different.
Note the similarity between Figure
\ref{figresonances}(a) and the scar function of Figure
\ref{figfscars2}(a). A single element of our base captures almost
all the structure of an exact quantum eigenstate. On the other hand,
the resonance shown in Figure \ref{figresonances}(b) results
essentially from the combination of the scar function in Figure
\ref{figfscars1}(a) and the one in Figure \ref{figfscars2}(b).

We postpone a more detailed analysis to a future publication, but
the convenience of our approach to the study of the phase space
morphology of quantum resonances is clear. Indeed, scar functions
have proved extremely useful in the study of scarring effects,
providing for example ways to quantify scarring and to understand
the influence of homoclinic motion on scars \cite{clinic}. We expect
it will also permit a more systematic study of scarring effects in
open systems \cite{pre77dw2008}, a subject still in its infancy. For
instance, there is certainly an interesting interplay between
scarring and the decay rate, so that more scarred states are
expected to live longer. We believe our approach will shed some
light on this issue.

To conclude, we have introduced a theory based on short periodic
orbits for quantum chaotic scattering. It may be argued that the use
of the scar matrix offers no real advantage since its construction
makes use of the quantum propagator. However, previous studies of
closed systems \cite{stadium} suggest that semiclassical approaches
can be successfully implemented within this framework, because the
propagation times involved are not longer than $T_E$. We are
currently working in this direction. Another direction to follow is
to adapt the theory to dielectric boundary conditions in order to
treat microlasers, an application where spectacular manifestations
of scarring can be observed \cite{lasers}.

We acknowledge partial support by EPSRC, the Royal Society, Fapesp,
CONICET, ANPCyT, and UBACyT.

\end{document}